\documentclass[journal]{IEEEtran}

\usepackage{booktabs}
\usepackage{bm}
\usepackage{graphicx}
\usepackage[cmex10]{amsmath}
\usepackage{cite}
\usepackage{amssymb}
\usepackage{subfigure}
\usepackage{extarrows}
\usepackage[letterspace = -2]{microtype}
\usepackage{algorithm}
\usepackage{algpseudocode}
\usepackage{algpascal}
\usepackage{float}
\newfloat{algorithm}{t}{lop}

\begin{document}

\title{
Joint Wireless Information and Power Transfer for an Autonomous Multiple Antenna Relay System
}

\author{
Yang~Huang,~\IEEEmembership{Student Member,~IEEE},
and Bruno~Clerckx,~\IEEEmembership{Member,~IEEE}

\thanks{This work has been partially supported by the EPSRC of UK under grant EP/M008193/1.}
\thanks{Y. Huang and B. Clerckx are with the Department of Electrical and Electronic Engineering, Imperial College London, London SW7 2AZ, United Kingdom (e-mail: \{y.huang13, b.clerckx\}@imperial.ac.uk). B. Clerckx is also with the School of Electrical Engineering, Korea University, Korea.}

}

\maketitle
\begin{abstract}
Considering a three-node multiple antenna relay system, this paper proposes a two-phase amplify-and-forward (AF) relaying protocol, which enables the autonomous relay to simultaneously harvest wireless power from the source information signal and from an energy signal conveyed by the destination. We first study this energy-flow-assisted (EFA) relaying in a single-input single-output (SISO) relay system and aim at maximizing the rate. By transforming the optimization problem into an equivalent convex form, a global optimum can be found. We then extend the protocol to a multiple antenna relay system. The relay processing matrix is optimized to maximize the rate. The optimization problem can be efficiently solved by eigenvalue decomposition, after linear algebra manipulation. It is observed that the benefits of the energy flow are interestingly shown only in the multiple antenna case, and it is revealed that the received information signal and the energy leakage at the relay can be nearly separated by making use of the signal space, such that the desired signal can be amplified with a larger coefficient.
\end{abstract}


\section{Introduction}
During the uplink transmission in a sensor network, the relay may suffer from energy drain, while the destination, as a central processing unit, has access to a reliable source of energy. Motivated by this scenario, a joint wireless information and power transfer (JWIPT) \cite{ZH13} is investigated in an autonomous relay network, where the relay harvests energy from the incoming signal from the source to forward information to the destination but is also aided by a direct power transfer from the destination. The forms of wireless power harvesting and information relaying applied to current works on autonomous relays mainly focus on the power splitting (PS) relaying and the time switching (TS) relaying \cite{NZDK13, JZ14arXiv, DPEP13, NNZKD14arXiv}. Ref. \cite{NZDK13} proposed a PS relaying (where the relay extracts power for forwarding from the source information signal) and a TS relaying (where the relay harvests power from an energy signal sent by the source and then relays source information in a time-division manner). Another TS relaying, where the energy signal is sent by the destination, was studied in\cite{JZ14arXiv}. The PS relaying was also studied in multi-pair one-way relay networks \cite{DPEP13} and relay interference channels\cite{NNZKD14arXiv}. In these works, the PS relaying reduces the received information power at the relay. The TS relaying consumes more timeslots, though the wireless power can be harvested in a dedicated timeslot. Thus, these two methods may degrade the rate performance. To harvest sufficient power without consuming more timeslots, this paper proposes an energy-flow-assisted (EFA) two-phase amplify-and-forward (AF) one-way relaying protocol.

In the proposed protocol, by exploiting PS, the relay can harvest power from both the source information signal and a concurrent energy flow (EF) at the destination (in the form of a WPT), as illustrated in Fig. \ref{FigRelayingSchemes}(a). We first study the protocol for a single-input single-output (SISO) relay system and aim at maximizing the rate. By transforming the original nonconvex optimization problem into an equivalent convex form, a global optimum can be achieved. We then extend the protocol to a multiple antenna relay system. The relay processing matrix is optimized to maximize the rate. By linear algebra manipulation, the original problem is transformed into a generalized Rayleigh quotient problem which can be efficiently solved by eigenvalue decomposition. Simulation results reveal that only in the multiple antenna case, can the EFA relaying significantly outperform rate-wise the relaying without EF. It is shown that although the information signal to be forwarded is corrupted by the EF, the two signals can be nearly aligned with two orthogonal directions by manipulating the signal space, and the information signal can be amplified with a larger coefficient.

The remainder of this paper is organized as follows. The system model is formulated in Section \ref{SecSystemModel}. Section \ref{SecRelayingProtocols} then proposes the relaying protocols for the SISO and the multiple antenna relay systems. Section \ref{SecSimulationResults} discusses the simulation results. Finally, conclusions are drawn in Section \ref{SecConclusion}. Notations: Throughout the paper, matrices and vectors are in bold capital and bold lower cases, respectively. The notations $(\cdot)^T$, $(\cdot)^\ast$, $(\cdot)^H$, $\text{Tr}\{\cdot\}$, $\text{vec}(\cdot)$, $\|\cdot\|_F$, $\otimes$, and $\langle\cdot,\cdot\rangle$ represent the transpose, conjugate, conjugate transpose, trace, vectorization, Frobenius norm, Kronecker product, and the inner product, respectively.

\section{System Model}
\label{SecSystemModel}
\begin{figure}[!t]
\centering
\includegraphics[width = 2.3in]{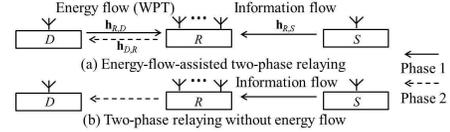}
\caption{JWIPT relay network.}
\label{FigRelayingSchemes}
\end{figure}
Fig. \ref{FigRelayingSchemes}(a) shows the EFA two-phase relaying, where the single-antenna source $S$ transmits information to the single-antenna destination $D$ via a $r$-antenna wireless-powered half-duplex AF relay $R$, without the direct link between $S$ and $D$. In phase 1, $D$ transmits an EF (i.e. an energy signal) in the form of a power transfer to $R$, in order to enhance the amount of harvested power at the relay. Simultaneously, $S$ transmits an information signal (i.e. an information flow) to $R$ for forwarding. The relay $R$ harvests power from both the EF and the information flow. In phase 2, the received signal at $R$ is processed and forwarded thanks to the wireless power harvested in phase 1. The $D$-$R$, $S$-$R$, and $R$-$D$ channels are respectively designated as $\mathbf{h}_{R,D} \in \mathbb{C}^{r \times 1}$, $\mathbf{h}_{R,S} \in \mathbb{C}^{r \times 1}$, and $\mathbf{h}_{D,R} \in \mathbb{C}^{r \times 1}$, which are independent and identically distributed (i.i.d.) Rayleigh flat fading channels. Channel reciprocity is assumed such that $\mathbf{h}_{D,R} = \mathbf{h}_{R,D}^T$. Global channel state information (CSI) is supposed to be perfectly obtained at $R$. In phase 1, the relay exploits PS for simultaneous energy harvesting (EH) and information detecting (ID). Specifically, at each antenna of the relay, a fraction of the received power, denoted as the PS ratio $\rho$ (where we consider the uniform PS), is conveyed to the EH receiver. The noise at the ID receiver at $R$ and the receiver at $D$ are respectively denoted by $\mathbf{n}_R \sim \mathcal{CN}(0, \sigma_n^2\mathbf{I})$ and $n_D \sim \mathcal{CN}(0, \sigma_n^2)$, while the noise at the EH receiver at $R$ is small and neglected.

In phase 1, the received signal at the EH receiver is given by $\mathbf{y}_{R,\text{EH}} = \rho^{1/2}(\mathbf{h}_{R, D} x_D + \mathbf{h}_{R, S} x_S)$, where $x_D$ and $x_S$ respectively denote the EF from $D$ and the information flow from $S$, and $\mathcal{E}\{|x_D|^2\} = P_D$ and $\mathcal{E}\{|x_S|^2\} = P_S$, where $P_D$ and $P_S$ denote the power budgets at $D$ and $S$. Assuming an RF-to-DC conversion efficiency of 1, the harvested wireless power at the EH receiver of $R$ is given by
\begin{equation}
\label{EqPwrRxSigEH}
P_R = \mathcal{E}\{\|\mathbf{y}_{R,\text{EH}}\|^2\} = \rho(\|\mathbf{h}_{R,D}\|^2P_D + \|\mathbf{h}_{R,S}\|^2P_S)\,.
\end{equation}
Meanwhile, the received signal at the ID receiver is given by $\mathbf{y}_{R,\text{ID}} = (1 - \rho)^{1/2}(\mathbf{h}_{R, S} x_S + \mathbf{h}_{R, D} x_D) + \mathbf{n}_R$. It shows that a part of the received EF (i.e. $(1 - \rho)^{1/2}\mathbf{h}_{R, D} x_D$), referred to as an energy leakage, leaks into the ID receiver and contaminates the information signal. This leakage cannot be removed due to AF. In phase 2, $\mathbf{y}_{R,\text{ID}}$ is processed by the relay processing matrix $\mathbf{F}$ and forwarded from $R$:
\begin{equation}
\label{EqTxSigR}
\mathbf{x}_{R,\text{ID}} = \mathbf{F}(1 - \rho)^{1/2}(\mathbf{h}_{R,S}x_S + \mathbf{h}_{R,D}x_D) + \mathbf{F}\mathbf{n}_R\,.
\end{equation}
The signal received at $D$ is given by $y_D^{\prime} = (1 - \rho)^{1/2} \mathbf{h}_{D, R}^T  \mathbf{F} \mathbf{h}_{R, S} x_S + (1 - \rho)^{1/2}\mathbf{h}_{D, R}^T \mathbf{F} \mathbf{h}_{R, D} x_D + \mathbf{h}_{D, R}^T\mathbf{F}\mathbf{n}_R + n_D$. Perfect knowledge of CSI and $\mathbf{F}$ at $D$ is assumed. Thus, the self-interference (i.e. the term containing $x_D$) is canceled, yielding
\begin{equation}
\label{EqRxSigD}
y_D = \mathbf{h}_{D,R}^T \mathbf{F} (1 - \rho)^{1/2} \mathbf{h}_{R,S}x_S + \mathbf{h}_{D,R}^T \mathbf{F} \mathbf{n}_R + n_D\,.
\end{equation}

When $P_D = 0$ (i.e. $x_D = 0$) as shown in Fig. \ref{FigRelayingSchemes}(b), the forwarding power only comes from the information flow by PS. This relaying protocol is referred to as the relaying without EF. Note that although more power is available to the relay for forwarding in the EFA relaying, a part of the power used for forwarding is consumed to retransmit the energy leakage. In the following sections, we study the above two relaying protocols in both SISO and multiple antenna relay systems.

\section{Relaying Protocols}
\label{SecRelayingProtocols}
\subsection{The SISO System}
In the SISO case, each node of $D$, $R$, and $S$ is equipped with one antenna, such that the $D$-$R$, $S$-$R$, and $R$-$D$ channels are respectively designated as $h_{R,D}$, $h_{R,S}$, and $h_{D,R}$. The relay processing matrix $\mathbf{F}$ also reduces to an amplification coefficient $f$. From (\ref{EqRxSigD}), the achievable rate is obtained as
\begin{equation}
\label{EqMutualInformation}
C_1 = 1/2\cdot\log_2 (1 + \gamma_1)\,,
\end{equation}
where the coefficient $1/2$ results from the half-duplex transmission, and the signal-to-noise ratio (SNR) $\gamma_1$ is given by
\begin{equation}
\label{EqRateSISO}
\gamma_1 = \frac{(1 - \rho)|h_{D,R}|^2 |f|^2 |h_{R,S}|^2 P_S}{(1 + |h_{D,R}|^2 |f|^2)\sigma_n^2} \,,
\end{equation}
where $|f|^2$ and $\rho$ are variables to be optimized. Since the power of the amplified signal at $R$ is no greater than the harvested wireless power, the power constraint at $R$ is given by
\begin{equation}
\label{EqOrigRelayPwrConstSISO}
(1 - \rho) p_1 |f|^2 + \sigma_n^2 |f|^2 \leq p_1 \rho \,,
\end{equation}
where $p_1 = |h_{R,S}|^2 P_S + |h_{R,D}|^2 P_D$. It is found in (\ref{EqRateSISO}) that $C_1$ increases as $|f|^2$ increases. When $|f|^2$ is maximized (i.e. $C_1$ is maximized), equality holds in (\ref{EqOrigRelayPwrConstSISO}). Thus,
\begin{equation}
\label{EqfsquareRelayPwrConstSISO}
|f|^2 = \rho p_1/((1- \rho) p_1 + \sigma_n^2)\,.
\end{equation}
The objective of the EFA relaying is maximizing the rate (\ref{EqMutualInformation}). Since $C_1$ is monotonically increasing, maximizing $C_1$ is equivalent to maximizing $\gamma_1$. Substituting (\ref{EqfsquareRelayPwrConstSISO}) into (\ref{EqRateSISO}), the optimization problem for the EFA relaying is formulated as
\begin{IEEEeqnarray}{cl}
\label{EqOriginalSisoSNRproblem}
\max_{\rho} \quad & \frac{\rho(1 - \rho)}{(1 - \rho)p_1 + \sigma_n^2 + |h_{D,R}|^2 p_1 \rho} \IEEEyessubnumber\label{EqOriginalSISOobj}\\
\text{s.t.} & 0 \leq \rho \leq 1\,.\IEEEyessubnumber
\end{IEEEeqnarray}
Problem (\ref{EqOriginalSisoSNRproblem}) is non-convex. However, the numerator of (\ref{EqOriginalSISOobj}), designated as $\phi(\rho)$, is concave and nonnegative, while the denominator, designated as $\psi(\rho)$, is convex and positive. Thus, (\ref{EqOriginalSisoSNRproblem}) is a concave fractional programming. By introducing variables $s \triangleq \rho/\psi(\rho)$ and $t \triangleq 1/\psi(\rho)$, (\ref{EqOriginalSisoSNRproblem}) can be equivalently transformed into a convex problem given by
\begin{IEEEeqnarray}{cl}
\label{EqConvexSisoSNRproblem}
\min_{s, t} \quad & - t \phi(s/t) \IEEEyessubnumber\\
\text{s.t.} & t \psi(s/t) \leq 1\,,\IEEEyessubnumber\\
& t > 0,\, 0\leq s\leq t\,,\IEEEyessubnumber
\end{IEEEeqnarray}
which can be solved by the interior-point algorithm\cite{BV04}. Then, the optimal $\rho^\star$ is obtained by $\rho^\star = s^\star/t^\star$, and the achievable rate (\ref{EqMutualInformation}) can be calculated with (\ref{EqfsquareRelayPwrConstSISO}) and (\ref{EqRateSISO}).

When $P_D = 0$, problem (\ref{EqConvexSisoSNRproblem}) boils down to the design problem of the relaying without EF. The optimal PS ratio can still be obtained by solving (\ref{EqConvexSisoSNRproblem}).

\subsection{The Multiple Antenna Relay System}
In the multiple antenna relay system (where $r > 1$), the system model is discussed in Section \ref{SecSystemModel}. According to (\ref{EqTxSigR}) and (\ref{EqRxSigD}), the power of the forwarded signal and the SNR $\gamma_2$ of the received signal at $D$ are respectively given by
\begin{IEEEeqnarray}{rcl}
\label{EqPwrTxSigR}
\mathcal{E}\{\|\mathbf{x}_{R, \text{ID}}\|^2\} &{}={}& (1 - \rho)\text{Tr}\left\{\mathbf{F}^H\mathbf{F} \left(\mathbf{h}_{R,S}\mathbf{h}^H_{R,S} P_S \right.\right.\nonumber\\ &&+\left.\left. \mathbf{h}_{R,D}\mathbf{h}_{R,D}^H P_D \right)\right\} + \text{Tr}\left\{\sigma_n^2 \mathbf{F}^H\mathbf{F}\right\}
\end{IEEEeqnarray}
and
\begin{equation}
\label{EqMultiAntennaOriginalSNR}
\gamma_2 = \frac{(1 - \rho)P_S|\mathbf{h}_{D,R}^T\mathbf{F}\mathbf{h}_{R,S}|^2}{\mathbf{h}_{D,R}^T\mathbf{F}\mathbf{F}^H\mathbf{h}_{D,R}^\ast\sigma_n^2 + \sigma_n^2}\,.
\end{equation}
Making use of $\text{Tr}\{\mathbf{A}\mathbf{B}\mathbf{C}\} = \text{vec} (\mathbf{A}^H)^H (\mathbf{C}^T\otimes\mathbf{I})\text{vec}(\mathbf{B})$, (\ref{EqPwrTxSigR}) can be written as
\begin{IEEEeqnarray}{rcl}
\label{EqPwrTxSigR_VecF}
\mathcal{E}\{\|\mathbf{x}_{R, \text{ID}}\|^2\}&=&\text{vec}(\mathbf{F})^H \! \left( \! (1 - \rho)\mathbf{Q}_R^T \! \otimes \! \mathbf{I} \! + \! \sigma_n^2 \mathbf{I} \right) \! \text{vec}(\mathbf{F}),
\end{IEEEeqnarray}
where $\mathbf{Q}_R \! = \! \mathbf{h}_{R,S}\mathbf{h}_{R,S}^H P_S + \mathbf{h}_{R,D}\mathbf{h}_{R,D}^H P_D$. Similarly, thanks to $\text{vec}(\mathbf{A}\mathbf{B}\mathbf{C}) \! = \! (\mathbf{C}^T \! \otimes \! \mathbf{A})\text{vec}(\mathbf{B})$, terms $\mathbf{h}_{D,R}^T \mathbf{F}\mathbf{h}_{R,S}$ and $\mathbf{h}_{D,R}^T  \mathbf{F}\mathbf{F}^H\mathbf{h}_{D,R}^\ast$ in (\ref{EqMultiAntennaOriginalSNR}) can be respectively rewritten as $\mathbf{h}_{D,R}^T \mathbf{F} \mathbf{h}_{R,S} \! = \! \text{vec}(\mathbf{h}_{D,R}^T \mathbf{F} \mathbf{h}_{R,S}) \! = \! (\mathbf{h}_{R,S}^T \otimes \mathbf{h}_{D,R}^T) \text{vec}(\mathbf{F})$ and $\mathbf{h}_{D,R}^T \mathbf{F} \mathbf{F}^H \mathbf{h}_{D,R}^\ast \! = \! \text{Tr}\{\mathbf{F}^H \mathbf{h}_{D,R}^\ast \mathbf{h}_{D,R}^T \mathbf{F}\} = \text{vec}(\mathbf{F})^H (\mathbf{I} \otimes \mathbf{h}_{D,R}^\ast \mathbf{h}_{D,R}^T) \text{vec}(\mathbf{F})$.
By defining $\mathbf{f} \triangleq \text{vec}(\mathbf{F})$, $\mathbf{\tilde{Q}_R} \triangleq \mathbf{Q}_R^T \otimes \mathbf{I}$, $\mathbf{K} \! \triangleq \! (\mathbf{h}_{R,S}^T \otimes \mathbf{h}_{D,R}^T)^H(\mathbf{h}_{R,S}^T \otimes \mathbf{h}_{D,R}^T) = [(\mathbf{h}_{R,S}\mathbf{h}_{R,S}^H) \otimes (\mathbf{h}_{D,R} \cdot \mathbf{h}_{D,R}^H)]^T$, and $\mathbf{J} \triangleq \mathbf{I} \otimes (\mathbf{h}_{D,R}^\ast\mathbf{h}_{D,R}^T) = [\mathbf{I} \otimes (\mathbf{h}_{D,R} \mathbf{h}_{D,R}^H)]^T$, (\ref{EqMultiAntennaOriginalSNR}) can be rewritten as
\begin{equation}
\label{EqMultiAntennaReformulatedSNR}
\gamma_2 = P_S \mathbf{f}^H \mathbf{\tilde{K}} \mathbf{f}/\left(\sigma_n^2 \left( \mathbf{f}^H\mathbf{J}\mathbf{f} + 1 \right)\right)\,,
\end{equation}
where $\mathbf{\tilde{K}} \triangleq (1 - \rho)\mathbf{K}$. The rate of the multiple antenna relay system is obtained by $C_2 = 1/2\cdot\log_2 (1 + \gamma_2)$.

Similarly to the SISO system, maximizing the rate $C_2 $ is equivalent to maximizing $\gamma_2$. It is shown in the subsequent discussion that the coupled $\rho$ and $\mathbf{f}$ are actually subject to an equality constraint (as shown in (\ref{EqRelayPwrConstMultiAntennaSNRproblem})), such that they cannot be optimized (or updated) alternatively. Therefore, as an initial study, in this SNR maximization problem, only $\mathbf{f}$ (i.e. the vectorization of the relay matrix) is optimized for simplicity. The best PS ratio $\rho$ is exhaustively searched over all PS ratios to maximize the SNR in (\ref{EqMultiAntennaReformulatedSNR}). The power constraint at $R$ is such that the transmit power (\ref{EqPwrTxSigR_VecF}) is no greater than the power budget (\ref{EqPwrRxSigEH}). Namely,
\begin{equation}
\label{EqOriginalPwrConstRMultiantenna}
(1 - \rho)\mathbf{f}^H\mathbf{\tilde{Q}}_R\mathbf{f} + \sigma_n^2\mathbf{f}^H\mathbf{f} \leq P_R\,.
\end{equation}
By decomposing $\mathbf{f}$ as $\mathbf{f} = \mathbf{\tilde{f}} \|\mathbf{f}\|$, where $\mathbf{\tilde{f}}$ denotes the normalized direction vector of $\mathbf{f}$, we find that the SNR $\gamma_2$ in (\ref{EqMultiAntennaReformulatedSNR}) increases as $\|\mathbf{f}\|^2$ increases. Hence, when (\ref{EqMultiAntennaReformulatedSNR}) is maximized, equality holds in (\ref{EqOriginalPwrConstRMultiantenna}); otherwise, a larger $\|\mathbf{f}\|^2$ can be found. Therefore, (\ref{EqOriginalPwrConstRMultiantenna}) can be rewritten as $((1 - \rho)\mathbf{f}^H\mathbf{\tilde{Q}}_R\mathbf{f} + \sigma_n^2\mathbf{f}^H\mathbf{f})/P_R = 1$. With the above equation, the term $\mathbf{f}^H\mathbf{J}\mathbf{f} + 1$ in the denominator of (\ref{EqMultiAntennaReformulatedSNR}) can be rewritten as $\mathbf{f}^H\mathbf{\tilde{J}}\mathbf{f}$, where $\mathbf{\tilde{J}} \triangleq \mathbf{J} + ((1 - \rho)\mathbf{\tilde{Q}}_R + \mathbf{I}\sigma_n^2)/P_R$. Hence, the SNR maximization problem can be formulated as
\begin{IEEEeqnarray}{ll}
\label{EqMultiAntennaSNRproblem}
\max_{\mathbf{f}} \,&\, \gamma_2^{\prime} = \mathbf{f}^H \mathbf{\tilde{K}} \mathbf{f}/\left(\mathbf{f}^H\mathbf{\tilde{J}}\mathbf{f}\right)\IEEEyessubnumber \label{EqObjMultiAntennaSNRproblem} \\
\text{s.t.} \, &\,(1 - \rho)\mathbf{f}^H\mathbf{\tilde{Q}}_R\mathbf{f} + \sigma_n^2\mathbf{f}^H\mathbf{f} = P_R \IEEEyessubnumber \label{EqRelayPwrConstMultiAntennaSNRproblem}\,,
\end{IEEEeqnarray}
where (\ref{EqObjMultiAntennaSNRproblem}) is a generalized Rayleigh quotient which can then be reduced to a Rayleigh quotient by defining $\mathbf{f} \triangleq \mathbf{L}_J^{-1}\mathbf{g} = \mathbf{L}_J^{-1}\mathbf{\tilde{g}}\|\mathbf{g}\|$, where $\mathbf{L}_J$ is achieved by the Cholesky decomposition $\mathbf{\tilde{J}} = \mathbf{L}_J^H\mathbf{L}_J$ and $\mathbf{\tilde{g}}$ is the direction vector of $\mathbf{g}$ such that $\mathbf{\tilde{g}}^H \mathbf{\tilde{g}} = 1$. Thus, problem (\ref{EqMultiAntennaSNRproblem}) can be transformed into an equivalent form
\begin{IEEEeqnarray}{ll}
\label{EqMultiAntennaSNRproblemWRTg}
\max_{\mathbf{\tilde{g}}, \|\mathbf{g}\|} &\quad \mathbf{\tilde{g}}^H \left(\mathbf{L}_J^{-1}\right)^H \mathbf{\tilde{K}}\mathbf{L}_J^{-1} \mathbf{\tilde{g}} \IEEEyessubnumber \label{EqObjFuncWRTgMultiAntennaSNRProblem} \\
\text{s.t.} & \left\|\mathbf{g}\right\|^2 \! \mathbf{\tilde{g}}^H \! \left[\mathbf{L}_J^{-1}\right]^H \! \left(\! (1\! -\! \rho) \mathbf{\tilde{Q}}_R \! + \! \sigma_n^2 \mathbf{I} \! \right)\! \mathbf{L}_J^{-1} \! \mathbf{\tilde{g}} \! = \! P_R. \IEEEyessubnumber
\label{EqRelayPwrConstWRTgMultiAntennaSNRproblem}
\end{IEEEeqnarray}
The optimal $\mathbf{\tilde{g}}^\star$ of problem (\ref{EqMultiAntennaSNRproblemWRTg}) is equal to the dominant eigenvector of $\left(\mathbf{L}_J^{-1}\right)^H \mathbf{\tilde{K}}\mathbf{L}_J^{-1}$, such that the maximum value of the objective function (\ref{EqObjFuncWRTgMultiAntennaSNRProblem}) is equal to the largest eigenvalue of $\left(\mathbf{L}_J^{-1}\right)^H \mathbf{\tilde{K}}\mathbf{L}_J^{-1}$. The optimum $\|\mathbf{g}^\star\|$ can be obtained by solving (\ref{EqRelayPwrConstWRTgMultiAntennaSNRproblem}) with the achieved $\mathbf{\tilde{g}}^\star$. Thereby, the rate can be calculated with (\ref{EqMultiAntennaReformulatedSNR}).

To gain an insight into the benefits of multiple antennas, we rewrite (\ref{EqObjMultiAntennaSNRproblem}) such that $\mathbf{f}$ is reshaped as $\mathbf{F}$. Given $\mathbf{\tilde{F}} = \mathbf{F} / \|\mathbf{F}\|_F$ and its singular value decomposition $\mathbf{\tilde{F}} = \mathbf{U} \mathbf{\Sigma} \mathbf{V}^H$ (where $\mathbf{\Sigma} = diag\{\lambda_1,\ldots,\lambda_r\}$ for $\lambda_1 \geq \cdots \geq \lambda_r$ and $\sum \lambda_i^2 = 1$), (\ref{EqObjMultiAntennaSNRproblem}) can be written as $\gamma_2^{\prime} = (1 - \rho) |\sum \mathbf{\tilde{h}}_{D,R}^T \mathbf{u}_i \mathbf{v}_i^H  \mathbf{\tilde{h}}_{R,S} \lambda_i|^2/( \frac{\sum |\mathbf{\tilde{h}}_{D,R}^T \mathbf{u}_i|^2 \lambda_i^2}{\|\mathbf{h}_{R,S}\|^2} \! + \! \frac{\epsilon_1 \sum |\mathbf{\tilde{h}}_{R,S}^H \mathbf{v}_i|^2 \lambda_i^2}{p_2} \! + \! \frac{\epsilon_2 \sum |\mathbf{\tilde{h}}_{R,D}^H \mathbf{v}_i|^2 \lambda_i^2}{p_2} \! + \! \frac{\sigma_n^2}{\rho p_2})$, where $\mathbf{\tilde{h}}_{p,q}$ denotes the normalized directions of $\mathbf{h}_{p,q}$; $\mathbf{u}_i$ and $\mathbf{v}_i$ denote the $i$\,th columns of $\mathbf{U}$ and $\mathbf{V}$, respectively; $\epsilon_1 = (1-\rho)P_S/(\rho \|\mathbf{h}_{D,R}\|^2)$; $\epsilon_2 = (1-\rho)P_D/(\rho \|\mathbf{h}_{R,S}\|^2)$; $p_2 = \|\mathbf{h}_{R,D}\|^2 P_D + \|\mathbf{h}_{R,S}\|^2 P_S$. Further, $\gamma_2^{\prime} = (1 - \rho) |\sum \xi_{1,i} e^{j\theta_{1,i}} \xi_{2,i} e^{j\theta_{2,i}} \lambda_i|^2/( \frac{\sum \xi_{1,i}^2 \lambda_i^2}{\|\mathbf{h}_{R,S}\|^2} \! + \! \frac{\epsilon_1 \sum \xi_{2,i}^2 \lambda_i^2}{p_2} \! + \! \frac{\epsilon_2 \sum \xi_{3,i}^2 \lambda_i^2}{p_2} \! + \! \frac{\sigma_n^2}{\rho p_2})$, where $\xi_{1,i} e^{j\theta_{1,i}} = \langle\mathbf{\tilde{h}}_{D,R}^\ast, \mathbf{u}_i\rangle$, $\xi_{2,i} e^{j\theta_{2,i}} = \langle\mathbf{\tilde{h}}_{R,S}, \mathbf{v}_i\rangle$, $\xi_{3,i} e^{j\theta_{3,i}} = \langle\mathbf{\tilde{h}}_{R,D}, \mathbf{v}_i\rangle$ such that $\sum \xi_{m,i}^2 = 1$ for $m = 1,2,3$. The angle $\theta_{1,i}$ and the value $\xi_{1,i}$ denote the pseudo-angle and the cosine of the Hermitian angle (which is the angle between two complex vectors in the real vector space) between $\mathbf{\tilde{h}}_{D,R}^\ast$ and $\mathbf{u}_i$, respectively\cite{Scharnhorst01}. Likewise, $\xi_{2,i}$, $\theta_{2,i}$, $\xi_{3,i}$, and $\theta_{3,i}$ denote corresponding pseudo-angles and cosines. Given fixed $\theta_{m,i}$, $\lambda_i$ and $\xi_{3,i}$, the value of $\gamma_2^{\prime}$ is increased, if the ordering of $\xi_{1,i}$ (and $\xi_{2,i}$), $i = 1, \ldots, r$, follows that of $\lambda_i$, e.g. $\xi_{1,1} > \xi_{1,2} > \cdots > \xi_{1,r}$ for $\lambda_1 > \lambda_2 > \cdots > \lambda_r$. This implies that $\mathbf{u}_1$ and $\mathbf{v}_1$ should be close to $\mathbf{\tilde{h}}_{D,R}^\ast$ and $\mathbf{\tilde{h}}_{R,S}$, respectively. Then, with fixed $\theta_{m,i}$, $\lambda_i$, $\xi_{1,i}$ and $\xi_{2,i}$, to increase $\gamma_2^{\prime}$, $\xi_{3,r}$ should be as large as possible, i.e. $\mathbf{v}_r$ is close to $\mathbf{\tilde{h}}_{R,D}$. This $\mathbf{v}_r$ can always be found by rotating $\mathbf{v}_r$ around $\mathbf{\tilde{h}}_{R,S}$ such that the angle between $\mathbf{v}_r$ and $\mathbf{\tilde{h}}_{R,S}$ does not change (i.e. $\xi_{2,r}$ is fixed). The above insight can also be verified by simulations.

\label{SecSimulationResults}
\begin{figure*}[t]
\begin{minipage}[t][2.7in][t]{.3\textwidth}
\centering
\subfigure[Rate as a function of power splitting ratio at a channel realization.]{
\label{FigMultiAntennaR_RateVsRho}
\includegraphics[width = 1.9in]
{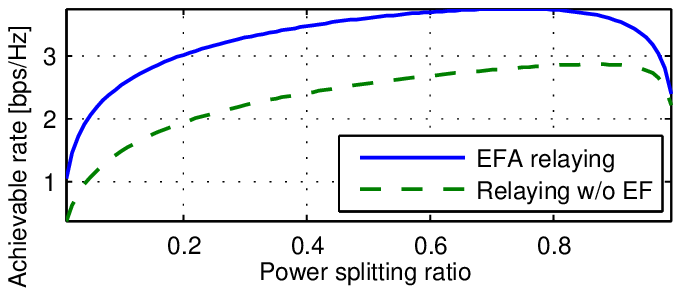}
}
\hfil
\subfigure[Average rate as a function of the number of antennas.]{
\label{Fig_MultiAntennaR_RateVsNumAntenna}
\includegraphics[width = 1.9in]
{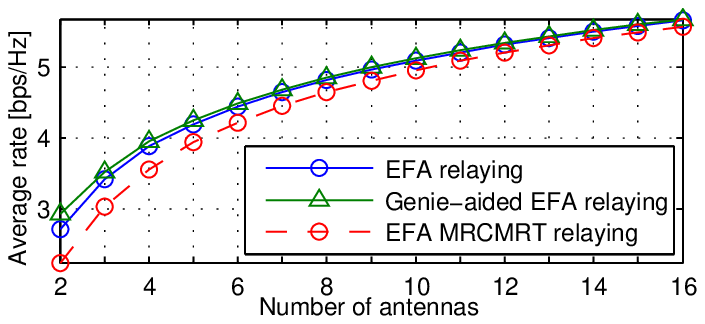}
}
\caption{The rate performance of the  multiple antenna EFA relaying with $d_{DR}/d_{DS} = 0.5$, $P_D = 0.5$\,W, and $P_S = 0.1$\,W.}\label{Fig_Rate_Performance_multiple_antenna}
\end{minipage}\hfill
\begin{minipage}[t][2.5in][t]{.28\textwidth}
\centering
\subfigure[$P_D = 0.5$\,W, $P_S = 0.1$\,W.]{
\label{Fig_SISO_Rate_AR2ABratio_dAB10_PA0p5_PB0p1}
\includegraphics[width = 2.0in]
{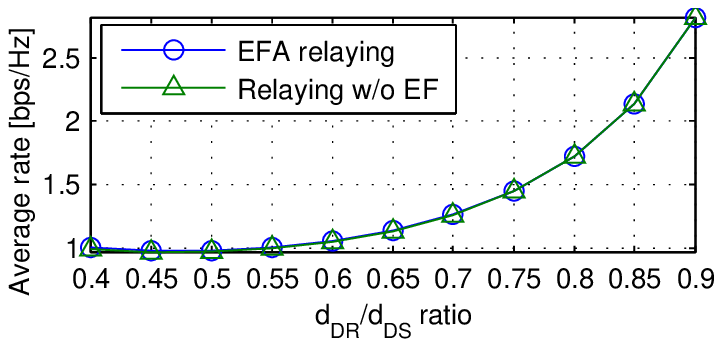}
}
\hfil
\subfigure[$P_D = 5$\,W, $P_S = 0.01$\,W.]{
\label{Fig_SISO_Rate_AR2ABratio_dAB10_PA5_PB0p01}
\includegraphics[width = 2.0in]
{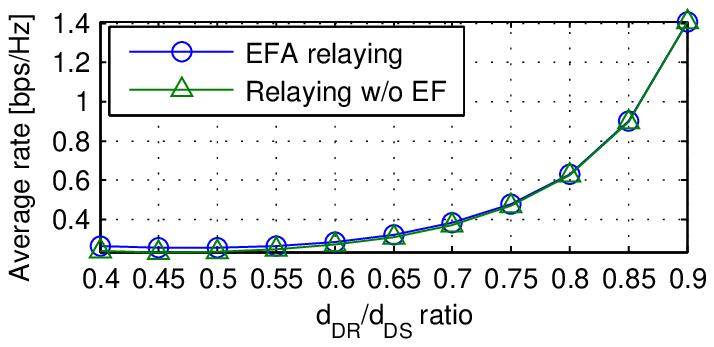}
}
\caption{Average rate as a function of $d_{DR}/d_{DS}$ in the SISO relay system.}\label{Fig_SISO_Performance_vs_AR2ABratio}
\end{minipage}\hfill
\begin{minipage}[t][2.5in][t]{.37\textwidth}
\centering
\subfigure[$P_D = 0.5$\,W, $P_S = 0.1$\,W.]{
\label{Fig_MultiAnteR_Rate_AR2ABratio_dAB10_PA0p5_PB0p1}
\includegraphics[width = 1.9in]
{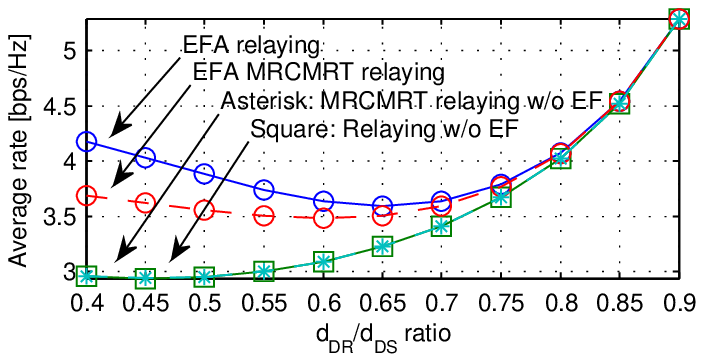}
}
\hfil
\subfigure[$P_D = 5$\,W, $P_S = 0.01$\,W.]{
\label{Fig_MultiAnteR_Rate_AR2ABratio_dAB10_PA5_PB0p01}
\includegraphics[width = 1.9in]
{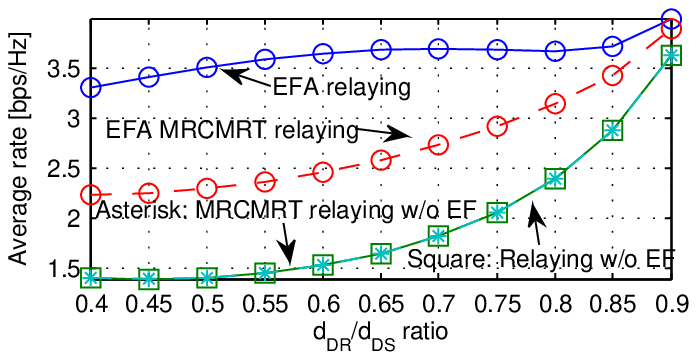}
}
\caption{Average rate as a function of $d_{DR}/d_{DS}$ in the multiple antenna relay system. The best PS ratio is exhaustively searched among 0.01:0.01:0.99.}\label{Fig_MultiAnteR_Performance_vs_AR2ABratio}
\end{minipage}\hfill
\end{figure*}

By exploiting maximum ratio combining (MRC) and maximum ratio transmission (MRT), a simplified relay matrix can be designed as $\mathbf{F}^{\prime} = \frac{\eta \mathbf{h}_{D,R}^\ast \mathbf{h}_{R,S}^H}{\|\mathbf{h}_{D,R}\| \|\mathbf{h}_{R,S}\|}$. The baseline schemes with $\mathbf{F}^{\prime}$ are referred to as EFA MRCMRT relaying and MRCMRT relaying without EF. The rate maximization problems (where $\eta$ and $\rho$ are the optimization variables) of these two MRCMRT protocols can be formulated as problems similar to (\ref{EqOriginalSisoSNRproblem}) and can also be equivalently transformed into convex forms similar to (\ref{EqConvexSisoSNRproblem}). The details are omitted here, due to the space constraint. It is noteworthy that in the MRCMRT relaying without EF, $\mathbf{F}^{\prime} = \arg \max_{\mathbf{F}} \gamma_2^{\prime}(\rho, \mathbf{F})$ \cite{KYA08}, which indicates that the MRCMRT structure is optimal for the scenario without EF. This is verified by Fig. \ref{Fig_MultiAnteR_Performance_vs_AR2ABratio}, where the rate of the MRCMRT relaying without EF coincides with that of the relaying without EF (with an exhaustive search of $\rho$).

\section{Simulation Results}
In the simulations, the channels are modeled as $\mathbf{h}_{p,q} = d_{pq}^{-3/2} \mathbf{\bar{h}}_{p,q}$, where $\mathbf{\bar{h}}_{p,q}$ denotes the small-scale fading; $d_{pq}$ is the distance between nodes $p$ and $q$, and $d_{DS} \triangleq d_{DR} + d_{RS}$. In the simulations, $d_{DS} = 10$\,m, $\sigma_n^2 = 1$\,$\mu$W, and $r = 4$ in the multiple antenna relay system. The power budget settings consider the symmetric case of $P_D = 0.5$\,W and $P_S = 0.1$\,W and the asymmetric case of $P_D = 5$\,W and $P_S = 0.01$\,W.

Fig. \ref{FigMultiAntennaR_RateVsRho} shows that the rate curves of the EFA relaying and the relaying without EF are concave. This is because a low PS ratio results in a limited forwarding power budget, while a high PS ratio reduces the receive SNR at $R$. Both the above two factors can lead to a limited SNR at $D$. In Fig. \ref{Fig_MultiAntennaR_RateVsNumAntenna}, the genie-aided EFA relaying refers to the EFA protocol where the energy leakage is ideally eliminated in the forwarded signal, i.e. the term containing $x_D$ in (\ref{EqTxSigR}) is removed. Thus, no forwarding power is consumed for the energy leakage, and the rate of the genie-aided relaying is the upper bound, as shown in Fig. \ref{Fig_MultiAntennaR_RateVsNumAntenna}. Since the relay beamforming of the EFA MRCMRT relaying does not address the energy leakage amplification problem, there is a gap between the rate of the EFA relaying and the EFA MRCMRT relaying. It is also observed that the rate of the EFA relaying can approach the ideal upper bound as the number of antennas increases.

As shown in Fig. \ref{Fig_SISO_Performance_vs_AR2ABratio}, the rate of the EFA relaying is negligibly higher than that of the relaying without EF in both the symmetric and the asymmetric cases. The reason lies in that the information signal and the energy leakage are amplified with the same amplification coefficient in the SISO relay, which means that the energy leakage may consume large amounts of the power for information forwarding. It is also observed that the rate of the EFA relaying nearly coincides with that of the relaying without EF when $R$ is close to $S$ in both the symmetric and asymmetric power budget cases. To figure out the reason, substituting (\ref{EqfsquareRelayPwrConstSISO}) into (\ref{EqRateSISO}) yields $\gamma_1 = (1 - \rho)\rho |h_{D,R}|^2 |h_{R,S}|^2 P_S/ \sigma_n^2/(1 - \rho + \sigma_n^2/p_1 + |h_{D,R}|^2\rho)$. Thus, when $R$ is close to $S$, $|h_{R,D}|^2$ may become very small such that the effect of $P_D$ on $\gamma_1$ is negligible.

Fig. \ref{Fig_MultiAnteR_Performance_vs_AR2ABratio} illustrates that the rate of the EFA relaying can be much higher than that of the relaying without EF (and MRCMRT relaying without EF) in both the symmetric and asymmetric cases, since the forwarding power is enhanced by EF. The EFA relaying also outperforms rate-wise the EFA MRCMRT relaying. Different from the SISO system, the EFA relaying in multiple antenna system can utilize the signal space to project the main parts of the desired information and the energy leakage onto orthogonal directions such that the information signal to be forwarded can be amplified with a larger coefficient, while the EFA MRCMRT relaying does not suppress the amplification of the energy leakage. It is also observed that the gaps between the rate of the relaying protocols decrease as $R$ moves towards the source. This is because the power of the EF received at $R$ decreases.

\section{Conclusion}
\label{SecConclusion}
In this paper, we have proposed an EFA two-phase relaying protocol. Protocols and optimization problems are designed and solved for both the SISO and the multiple antenna relay system. Simulation results show that EF is beneficial to the EFA relaying only in the presence of multiple antennas, because the information signal and the energy leakage can be nearly aligned with two orthogonal directions such that the desired signal can be amplified with a larger coefficient. The robust design addressing the problems of imperfect CSI and imperfect self-interference cancelation could be a future work.

\bibliographystyle{IEEEtran}
\bibliography{IEEEabrv,BibPro}

\end{document}